\documentclass[epj,numbook]{svjour}
\pdfoutput=1

\usepackage{amsmath,bm,dsfont,graphicx,slashed}
\usepackage{flushend} 

\DeclareMathOperator\Tr{Tr}
\DeclareMathOperator\tr{tr}
\DeclareMathOperator\re{Re}
\DeclareMathOperator\diag{diag}

\newcommand{\D}{D}
\newcommand{\kakko}[1]{\left\langle{#1}\right\rangle}
\newcommand{\mkakko}[1]{\left({#1}\right)}
\newcommand\1{\mathds 1}
\newcommand\mE{\mathcal{E}}
\newcommand\mL{\mathcal{L}}
\newcommand\mO{\mathcal{O}}
\newcommand\del{\partial}
\newcommand{\SU}{\text{SU}}
\newcommand{\SO}{\text{SO}}
\newcommand{\Sp}{\text{Sp}}
\newcommand{\U}{\text{U}}
\newcommand{\sgn}{\text{sgn}}
\newcommand\eps{\varepsilon}
\renewcommand\epsilon{\varepsilon}
\renewcommand{\bar}[1]{\overline{#1}} 
\newcommand{\hh}{H_2}
\newcommand\bset[2]{\underset{\makebox[0mm]{\scriptsize$#1$}}{\underbrace{#2}}}
\newcommand{\ym}{\mathrm{YM}}
\renewcommand{\kappa}{\epsilon}

\begin{document}\sloppy

\title{Banks-Casher-type relation for the BCS gap at high
  density\thanks{Preprint INT-PUB-12-053, YITP-12-70}}

\author{Takuya Kanazawa\inst{1}
  \and Tilo Wettig\inst{2} \and Naoki Yamamoto\inst{3,4}}

\institute{Department of Physics, The University of Tokyo, 7-3-1
  Hongo, Bunkyo-ku, Tokyo 113-0033, Japan \and Department of Physics,
  University of Regensburg, 93040 Regensburg, Germany \and Yukawa
  Institute for Theoretical Physics, Kyoto University, Kyoto 606-8502,
  Japan \and Maryland Center for Fundamental Physics, Department of
  Physics, University of Maryland, College Park, MD 20742-4111, USA}

\date{Received: date / Revised version: date}

\abstract{
  We derive a new Banks-Casher-type relation which relates the density
  of complex Dirac eigenvalues at the origin to the BCS gap of quarks
  at high density and zero temperature.  Our relation is applicable to
  QCD and QCD-like theories without a sign problem, such as two-color
  QCD and adjoint QCD with baryon chemical potential, and QCD with
  isospin chemical potential.  It provides us with a method to measure
  the BCS gap through the Dirac spectrum on the lattice.
}

\maketitle

\allowdisplaybreaks[1]

\section{Introduction}

Symmetries and their spontaneous breaking are important ingredients
for a modern understanding of physics. In quantum chromodynamics
(QCD), the spontaneous breaking of chiral symmetry, characterized by
the chiral condensate $\langle \bar \psi \psi \rangle$, gives rise to
a variety of phenomena in low-energy hadron physics.  Understanding
the nature and the mechanism of chiral symmetry breaking is a
fundamental challenge for theoretical approaches to QCD.

More than 30 years ago, it was realized by Banks and Casher
\cite{Banks:1979yr} that in the QCD vacuum the Dirac eigenvalue
spectrum at the origin is related to the magnitude of the chiral
condensate in the chiral limit through the relation
\begin{equation}
  \label{eq:BC0}
  |\langle \bar \psi \psi \rangle| = \pi \rho(0)\,,
\end{equation}
where $\rho(\lambda)$ is the spectral density of the Dirac operator.
This relation allows us to measure the magnitude of the chiral
condensate through the Dirac eigenvalue spectrum in lattice
simulations \cite{Hands:1990wc,BerbenniBitsch:1997fj,Giusti:2008vb}.
Relation \eqref{eq:BC0} was derived for purely imaginary (or real,
depending on the convention) eigenvalues.  It is valid not only in the
QCD vacuum but also at nonzero temperature $T$, as $T$ does not
invalidate any part of the original derivation.  Since $\rho(0)$ is
nonzero in the phase with broken chiral symmetry, whereas it vanishes
in the high-temperature chirally symmetric phase, it serves as a clean
indicator of the chiral phase transition
\cite{Lin:2011wb,Ohno:2011yr,Cossu:2012gm}.  In this way the Dirac
eigenvalue spectrum plays a crucial role in the modern understanding
of the nonperturbative phenomena of chiral symmetry breaking and
restoration.  The Banks-Casher relation is also valid in QCD-like
theories, including two-color QCD and QCD with fermions in the adjoint
representation, which share some universal features with \emph{bona
  fide} QCD \cite{Leutwyler:1992yt,Smilga:1994tb,Kogut:2000ek}.

An interesting question to ask is whether the Banks-Casher relation
could be generalized to QCD and QCD-like theories at nonzero quark
chemical potential $\mu$, whose phase diagrams have been studied
extensively (see, e.g., \cite{Fukushima:2010bq,vonSmekal:2012vx} for
recent reviews).  For small $\mu$, there is no Banks-Casher relation
but a more complicated connection between the chiral condensate and the
spectrum of the Dirac operator \cite{Osborn:2005ss,Osborn:2008jp}.  At
large $\mu$, another interesting nonperturbative phenomenon is
expected to occur \cite{Alford:2007xm}: quark-quark pairs are formed
near the Fermi surface due to the Bardeen-Cooper-Schrieffer (BCS)
mechanism.  Connections between BCS pairing and the Dirac eigenvalue
spectrum were first investigated in \cite{Yamamoto:2009ey} for
three-color QCD and in
\cite{Kanazawa:2009ks,Kanazawa:2009en,Akemann:2010tv} for two-color
QCD.  Nevertheless, the Banks-Casher relation itself has not yet been
generalized to dense quark matter.\footnote{A Banks-Casher-type
  relation that relates the spectrum of the Dirac \emph{singular
    values} at the origin to the condensate of quark-quark pairs in
  QCD-like theories at \emph{any} $\mu$ has been obtained in
  \cite{Kanazawa:2011tt}.}

In this paper, we derive a new Banks-Casher-type relation in theories
without a sign problem, which states that the square of the BCS gap
(or the superfluid gap) of quarks at asymptotically high
density\footnote{Strictly speaking, the new Banks-Casher-type relation
  is not exact at intermediate density. For example, instanton
  effects, which are negligible at asymptotically high density, may
  quantitatively alter our relation at intermediate density (see
  footnote \ref{foot:het}). Still, instanton effects are suppressed by
  high powers of $\Lambda_{\rm QCD}/\mu$
  \cite{Schafer:2002ty,Yamamoto:2008zw}, and are expected to be minor
  for $\mu$ larger than $\Lambda_{\rm QCD}$.  Whether the Dirac
  spectrum for a given $\mu$ is related to the chiral condensate or
  the BCS gap can be judged by looking at the spacing of the near-zero
  eigenvalues as a function of the four-volume $V_4$: If it is
  proportional to $1/\sqrt{V_4}$ ($1/V_4$), the Dirac spectrum is
  related to the BCS gap (chiral condensate).  A deviation from the
  $1/\sqrt{V_4}$ scaling at intermediate density due to instanton
  effects is expected to become smaller as $\mu$ becomes large
  compared with $\Lambda_{\rm QCD}$.} and zero temperature in the
chiral limit is proportional to the average density of near-zero
complex eigenvalues of the non-Hermitian Dirac operator. The explicit
form of the relation is
\begin{equation}
  \label{eq:main}
  \Delta^2 = \frac{2\pi^3}{3d_\text{rep}} \rho_1(0)\,,
\end{equation}
where $\Delta$ is the energy gap in the spectrum of the quasiquarks
above the ground state, $\rho_1(\lambda)$ is the two-dimensional
spectral density, see \eqref{eq:one} for the definition, and
$d_\text{rep}$ is the dimension of the color representation in which
the quarks transform: $d_\text{fund}=N_c$ and $d_\text{adj}=N_c^2-1$
for the fundamental and adjoint representation of $\SU(N_c)$,
respectively.  We prove this relation for two-color QCD
\cite{Kogut1999,Kogut:2000ek} with baryon chemical potential, QCD with
isospin chemical potential \cite{Alford:1998sd,Son:2000xc,Son:2000by},
and QCD with fermions in the adjoint representation
\cite{Kogut:2000ek} with baryon chemical potential.\footnote{The Dirac
  operators in these theories belong to three different symmetry
  classes \cite{Verbaarschot:2000dy,Akemann:2007rf}.}  All of these
theories have a positive-definite path integral measure if the
parameters of the theory are chosen properly.  On the other hand, our
relation does not extend to color-superconducting phases in
three-color QCD with baryon chemical potential,\footnote{It is worth
  stressing that a link between the Dirac spectrum and the BCS pairing
  \emph{does exist} even for a color-superconducting phase
  \cite{Yamamoto:2009ey}.  However, in such a case $\Delta^2$ receives
  contributions from an extended region within the Dirac spectrum and
  is not simply proportional to $\rho_1(0)$.}  because the spectral
density of the Dirac operator near the origin seems to be well-defined
in the thermodynamic limit \emph{only} for theories with a
positive-definite path integral measure
\cite{Leutwyler:1992yt,Osborn:2005ss,Osborn:2008ab} (see also appendix
C of \cite{Kanazawa:2011tt} for a detailed discussion of this point).
Nevertheless, our new relation serves as a useful means of measuring
the magnitude of the BCS gap at high density in lattice simulations of
theories without a sign problem.  Together with the spectral sum rules
for Dirac eigenvalues in the microscopic limit \cite{Kanazawa:2009ks},
our result constitutes a unique and quantitative way of studying the
physics of relativistic dense matter.

This paper is organized as follows. In sections~\ref{sec:two-color},
\ref{sec:iso}, and \ref{sec:adj} we derive the Banks-Casher-type
relation for the BCS gap in two-color QCD with baryon chemical
potential, QCD with isospin chemical potential, and adjoint QCD with
baryon chemical potential, respectively.  Conclusions and outlook are
presented in section \ref{sec:summary}.  In appendix~\ref{app:h2} we
give the microscopic derivation of a high-energy constant appearing in
the effective Lagrangian, and in appendix~\ref{app:sumrules} we
discuss the consequences of a high-energy term that was missing in
\cite{Kanazawa:2009ks}.  In appendix~\ref{app:free} we derive the free
Dirac spectrum at nonzero $\mu$, while in appendix~\ref{app:replica}
we comment on an alternative derivation of the Banks-Casher-type
relation based on the replica method.

In this paper we always work in Euclidean spacetime and at zero
temperature.  Our convention is that the average of an observable $O$
over gauge fields weighted by the pure Yang-Mills action is denoted by
$\kakko{O}_{\ym}$, while the average including $n$ dynamical flavors
is denoted by $\kakko{O}_{N_f=n}$.  By definition,
$\kakko{O}_{\ym}=\kakko{O}_{N_f=0}$\,.

\section{Two-color QCD}
\label{sec:two-color}

In this section we consider two-color QCD at high density, i.e., for
sufficiently large quark chemical potential $\mu$ such that
$\Lambda_{\SU(2)}\ll\mu$, where $\Lambda_{\SU(2)}$ is the typical
scale of the strong interactions in two-color QCD.  In
section~\ref{sec:deriv} we derive a Banks-Casher-type relation between
the BCS gap squared and the Dirac eigenvalue density at the origin,
working with two flavors for simplicity.  In section~\ref{sec:cons} we
show that this result is consistent with earlier results
\cite{Akemann:2010tv} on the microscopic Dirac spectrum.  In
section~\ref{sec:sumrules} we derive consistency relations based on
known spectral sum rules \cite{Kanazawa:2009ks}.  In
section~\ref{sec:general} we show that our results are also valid for
general even $N_f$.

\subsection{Derivation of Banks-Casher-type relation}
\label{sec:deriv}

The partition function of two-color QCD at $\mu\ne0$ is given by
\begin{align}
  Z(\{m_f\}) = \bigg\langle \prod_{f=1}^{N_f} \det(\D+m_f)
  \bigg\rangle_{\ym}\,,
  \label{eq:Zdet}
\end{align}
where $\D=\D(\mu)=\gamma_\nu D_\nu+\mu \gamma_4$ is the Dirac operator
and the $m_f$ are the quark masses.  In the following we take $N_f=2$
for simplicity (comments on general $N_f$ will be given in
section~\ref{sec:general}).  The partition can then be written as
\begin{align} 
  Z(m_1,m_2)
  & = \big\langle{\det(\D+m_1)\det(\D+m_2)}\big\rangle_{\ym} \notag\\
  & = \big\langle{\det(\D+m_1)\det(\D^\dagger+m_2)}\big\rangle_{\ym}\,.
  \label{eq:Z_2} 
\end{align}
The last line follows from the anti-unitary symmetry
$[C\gamma_5\tau_2K,D]=0$ of the two-color Dirac operator
\cite{Leutwyler:1992yt} (with the charge conjugation operator $C$, the
generator $\tau_2$ of the color group, and the complex conjugation
operator $K$), which implies $\D^*=(C\gamma_5\tau_2)^{-1} \D \,
(C\gamma_5\tau_2)$.

To derive Banks-Casher-type relations it is mandatory that the
integration measure is positive definite, see appendix~C of
\cite{Kanazawa:2011tt}.  Looking at \eqref{eq:Z_2}, we have two
options: $m_1=z$ and $m_2=z^*$ with $z$ complex, or $m_1=m_2=m$ with
$m$ real.  In the following we choose the first option and comment on
the second one at the end of this section.  Before proceeding we note
that \eqref{eq:Z_2} with $m_1=z$ and $m_2=z^*$ is similar in spirit
(although not quite identical) to the ``Hermitization'' of a
non-Hermitian problem, see
\cite{Sommers:1988zn,Haake1992,Stephanov:1996ki,Janik:1996xm,Feinberg:1997dk,Splittorff:2006uu}
for other applications of this method.

Let us consider the quark mass dependence of the partition function
(or the shift of the free energy due to the quark mass) in two-color
QCD at high density.  Recall that there is no contribution of
$\mO(m^1)$ at high density where the chiral condensate is vanishing.
The lowest-order shift in the free energy is thus $\mO(m^2)$.  In
\cite{Kanazawa:2009ks} the contribution $\delta\mE$ of the BCS pairing
of quarks near the Fermi surface to the mass dependence of the
ground-state energy density was computed based on high-density
effective theory (HDET) with quark masses \cite{Schafer:2001za}.
However, this result needs a small modification to allow for a mass
matrix of the form $M=\diag(z,z^*)$.  To see this we first note that
usually one introduces complex quark masses by writing the mass term
as $M_{LR}\frac{1+\gamma_5}{2}+M_{RL}\frac{1-\gamma_5}{2}$ and then
letting $M_{LR}=M$ and $M_{RL}=M^\dagger$.  However, in \eqref{eq:Z_2}
we have chosen $M_{LR}=M_{RL}=M=\diag(m_1,m_2)$, where $m_1$ and $m_2$
could be complex.\footnote{See also footnote 2 of
  \cite{Splittorff:2003cu} for a discussion of this subtlety.}  In
this case the result of \cite{Kanazawa:2009ks} for $N_f=2$ becomes
\begin{align}
  \label{eq:dE2}
  \delta\mE=\min_A\left\{-\frac{3}{2\pi^2}\Delta^2 
  (A^2+A^{*2})\det M\right\},
\end{align}
where $A\in\U(1)$.  Note that for $\delta\mE$ to be real, $\det M$ has
to be real, which is the case for $M=\diag(z,z^*)$.  For this choice
of $M$ the RHS is minimized for $A=\pm1$.  In $\mO(m^2)$ there is also
a high-energy term\footnote{We follow the convention in
  \cite{Gasser:1984gg}.  Our high-energy term is $\tr(M^2)$ instead of
  the more conventional $\tr({M^{\dag}M})$ for the same reason as
  discussed before \eqref{eq:dE2}.  Note that $\tr(M^2)$ is real for
  $M=\diag(z,z^*)$.} $H_2 \tr(M_{RL}M_{LR})=H_2\tr(M^2)$ that was not
considered in \cite{Kanazawa:2009ks} so
that\footnote{\label{foot:het}The explicit breaking of the $\U_A(1)$
  symmetry due to instantons is ignored here, based on the fact that
  at asymptotically high density the QCD running coupling is small and
  instantons are screened.  In other words, $\U_A(1)$ is assumed to be
  \emph{spontaneously} broken by $\kakko{\psi\psi}\ne 0$.  If we were
  to take the explicit breaking of $\U_A(1)$ into account, yet another
  mass term of the form $c_\text{inst}\det M$ would appear in the free
  energy, making it difficult to extract $\Delta^2$ separately from
  $c_\text{inst}$.  Finally, note that the contribution from the
  Bedaque-Sch\"{a}fer term \cite{Bedaque:2001je}, which affects the
  ground-state energy density at $\mO(m^4)$, is also ignored here.}
\begin{align}
  \frac1{V_4}\log Z(z,z^*)
  &=\frac{3}{\pi^2}\Delta^2zz^* + \hh (z^2+z^{*2}) + \mO(|z|^3)\,,
  \label{eq:logZ}
\end{align}
where $V_4$ is the four-volume and $\hh$ is a high-energy constant
whose origin will be clarified in appendix~\ref{app:h2}.\footnote{Note
  that the $\hh$ term, which originates from the \emph{bulk} Fermi sea
  and the vacuum, does not appear in \eqref{eq:dE2} because HDET
  captures the physics only near the Fermi surface. This is why we
  have to compute $\hh$ separately from microscopic QCD, see appendix
  \ref{app:h2}.}  Denoting $\del=\del/\del z$ and $\bar\del=\del/\del
z^*$ we obtain
\begin{align}    
  \lim_{z\to0}\lim_{V_4\to\infty}\frac1{V_4}\,
  \bar\del\del\log Z(z,z^*)=\frac{3}{\pi^2}\Delta^2\,,
  \label{eq:LHS}
\end{align}
i.e., we can extract $\Delta^2$ by suitable derivatives with respect
to the complex quark masses.

We now make contact with the Dirac spectrum by writing $Z(z,z^*)$ in
terms of the eigenvalues $\lambda_n$ of the Dirac
operator.\footnote{In principle the Dirac operator can have exact zero
  modes with $\lambda_n=0$, but since at high density topology is
  suppressed we can neglect exact zero modes in this paper.}  Starting
from \eqref{eq:Z_2} we follow \cite[eq.~(56)]{Fyodorov1997} and
introduce a small regulator $\kappa>0$ (see
\cite{Sommers:1988zn,Haake1992,Janik:1996xm,Fyodorov1997,Efetov:1997zza}
for a similar trick) to obtain
\begin{align}
  Z(z,z^*;\kappa) =
  \bigg\langle{\prod_n\big[(\lambda_n+z)(\lambda_n^*+z^*)
    +\kappa^2 \big]}\bigg\rangle_{\ym}\,.
  \label{eq:Z_2_kappa}
\end{align}
The necessity of this regulator is not obvious at this point and will
be explained after \eqref{eq:rho_main}.  Here we only note that it
ensures that the factors in the product can never become zero.  From
\eqref{eq:Z_2_kappa} we derive
\begin{align}
  &\bar\del\del\log Z(z,z^*;\kappa) 
  =\bigg\langle{\sum_n \frac{\kappa^2}{(|\lambda_n+z|^2
      +\kappa^2)^2} \bigg\rangle}_{N_f=2}\notag\\
  &+ \bigg\langle{\sum_m 
    \frac{\lambda_m^*+z^*}{|\lambda_m+z|^2+\kappa^2}
    \sum_n 
    \frac{\lambda_n+z}{|\lambda_n+z|^2+\kappa^2}}\bigg\rangle_{N_f=2}
  \notag\\
  &- \bigg\langle{\sum_n 
  \frac{\lambda_n^*+z^*}{|\lambda_n+z|^2+\kappa^2}}\bigg\rangle_{N_f=2}
  \bigg\langle{\sum_n 
  \frac{\lambda_n+z}{|\lambda_n+z|^2+\kappa^2}}\bigg\rangle_{N_f=2} \,.
  \label{eq:deldelZ}
\end{align}
We now take the thermodynamic limit ($V_4\to\infty$) and then the
chiral limit ($z\to0$).  Introducing the one-point
function\footnote{\label{foot:conv}Note two small changes in
  conventions compared to \cite{Kanazawa:2009ks}.  There, the Dirac
  eigenvalues were defined to be $i\lambda_n$, and the one-point
  function was defined without the factor of $1/V_4$.}
\begin{align}
  \rho_1(\lambda) = \lim_{z\to 0}\lim_{V_4\to\infty}\frac1{V_4}
  \Big\langle{\sum_n \delta^2(\lambda-\lambda_n)}\Big\rangle_{N_f=2}
  \label{eq:one}
\end{align}
and the connected two-point function
\begin{alignat}{1}
  &\rho_2^c(\lambda,\lambda')=\lim_{z\to 0}\lim_{V_4\to\infty}\notag\\
  &\bigg[\frac1{V_4} 
  \Big\langle{\sum_m \delta^2(\lambda\!-\!\lambda_m)
    \sum_n \delta^2(\lambda'\!-\!\lambda_n)}\Big\rangle_{\!N_f=2}
  \!-\! V_4\rho_1(\lambda)\rho_1(\lambda')\bigg]
\end{alignat}
we obtain
\begin{alignat}{1}
  &\lim_{z\to 0}\lim_{V_4\to \infty}
  \frac{1}{V_4}\,\bar{\del}\del \log Z(z,z^*;\kappa) 
  = \int d^2\lambda\,
  \frac{\kappa^2}{(|\lambda|^2+\kappa^2)^2}
  \rho_1(\lambda) 
  \notag
  \\
  &\quad + 
  \int d^2\lambda \int d^2\lambda'\,
  \frac{\lambda^*}{|\lambda|^2+\kappa^2}
  \frac{\lambda'}{|\lambda'|^2+\kappa^2}\,
  {\rho}_2^c(\lambda,\lambda')\,.
  \label{eq:rhs}
\end{alignat}
From chiral symmetry we have
${\rho}^c_2(\lambda,\lambda')={\rho}^c_2(-\lambda,\lambda')
={\rho}^c_2(\lambda,-\lambda')={\rho}^c_2(-\lambda,-\lambda')$, and
therefore the second line vanishes.  Using the formula
\begin{align}
  \delta^2(z) = \frac{1}{\pi} \lim_{\alpha\to 0}
  \frac{\alpha^2}{(|z|^2+\alpha^2)^2}
\end{align}
for the delta function in the complex plane we arrive at
\begin{align}
  \lim_{\kappa\to 0} \lim_{z\to 0} \lim_{V_4\to \infty}
  \frac{1}{V_4}\,\bar{\del}\del \log Z(z,z^*;\kappa) 
  = \pi {\rho}_1(0)\,.
  \label{eq:Z_rho_main}
\end{align}
Comparing this to \eqref{eq:LHS} we obtain our main result
\begin{align}
  \Delta^2 = \frac{\pi^3}{3}\rho_1(0)\,,
  \label{eq:rho_main}
\end{align}
which is a Banks-Casher-type relation for the BCS gap squared in terms
of the Dirac eigenvalue density at zero.

Our derivation assumes the high-density BCS-type superfluid phase of
QCD.  At low density, the Dirac spectrum has also been analyzed by
means of low-energy effective theories at the mean field level
\cite{Toublan:1999hx,Verbaarschot:2005rj,Osborn:2008ab}.  These
analyses at low density, which complement our analysis at high
density, showed that the Dirac spectrum at small $\mu$ forms a band of
width $4F^2\mu^2/\Sigma$ and height $\Sigma^2/(4\pi\mu^2F^2)$, which
extends infinitely along the imaginary axis.\footnote{There are typos
  in the spectral density in \cite[eq.~(38)]{Toublan:1999hx} and
  \cite[eq.~(34)]{Osborn:2008ab}. They must be divided by $\pi$.} Here
$F$ and $\Sigma$ are the usual low-energy constants (pion decay
constant and chiral condensate) at $\mu=0$.

In the rest of this subsection we make a number of technical comments
that can be skipped in a first reading.  

First, one can show that the contribution of the large eigenvalues to
the first integral on the RHS of \eqref{eq:rhs} is suppressed if we
introduce an ultraviolet cutoff $\Lambda_\text{UV}$ and take the limit
$\kappa\to 0$ \emph{before} the limit $\Lambda_\text{UV}\to\infty$.
Also, even if we had not neglected exact zero modes, their
contribution to the integral would have been suppressed in the limit
$V_4\to\infty$ in analogy to the discussion in
\cite{Leutwyler:1992yt}.

Second, let us explain the necessity of the regulator.  Without it we
would not have generated the first term on the RHS of
\eqref{eq:deldelZ}, which eventually gives the desired result.
Instead, the other two terms on the RHS would have contained
singularities for $\lambda_n+z=0$. These singularities need
regularization.  Note also that the regulator cannot be removed after
the derivation of \eqref{eq:deldelZ} since the first term on the RHS
would then become
$\big\langle\sum_n\pi\delta^2(\lambda_n+z)\big\rangle_{N_f=2}$.  While
this looks like the desired delta function we have to remember that
for $\kappa=0$ the fermionic measure is zero for $\lambda_n+z=0$, see
\eqref{eq:Z_2_kappa}.  Consequently, at any \emph{finite} volume this
term is equal to zero for any $z$.  Taking the thermodynamic limit
does not make it nonzero.  In conclusion, we only obtain the desired
result by using a regulator and removing it at the very end of the
calculation.  This is in contrast to the derivation of the usual
Banks-Casher relation at $\mu=0$, where the factors of $\lambda_n+m$
in the fermionic measure are always nonzero since in this case
$\lambda_n$ is purely imaginary and $m$ is real.

Third, we comment on the possibility of choosing real instead of
complex conjugate masses.  With $m_1=m_2=m$ the RHS of \eqref{eq:logZ}
becomes $(3\Delta^2/\pi^2+2\hh)m^2$ so that we cannot extract
$\Delta^2$ alone by taking derivatives with respect to
$m$.\footnote{\label{foot:problem}At lower density, where $\U_A(1)$ is
  explicitly broken by the axial anomaly, a similar problem occurs
  even with complex quark masses.  In this case the coefficient
  $c_\text{inst}$ mentioned in footnote~\ref{foot:het} is nonzero so
  that derivatives with respect to $z$ and $z^*$ extract a linear
  combination of $\Delta^2$ and $c_\text{inst}$.  }  However, as an
alternative to \eqref{eq:LHS} we can compute
\begin{align}
  \lim_{m\to0}\lim_{V_4\to\infty}\lim_{m_{1,2}=m}\frac1{V_4}\,
  \frac{\del^2}{\del m_1\del m_2}\log Z(m_1,m_2)
  =\frac{3}{\pi^2}\Delta^2 
\end{align}
to extract $\Delta^2$.  In the calculation of the other side of the
equation we first compute $\del_{m_1}\del_{m_2}\log Z(m_1,m_2;\kappa)$
at finite volume and then set $m_1=m_2=m$.  After this step, $z$ on
the RHS of \eqref{eq:deldelZ} is replaced by $m$ and the measure is
again positive definite.  The rest of the calculation then goes
through in the same way and yields \eqref{eq:rho_main}.  Nevertheless,
we have chosen to work with complex conjugate masses since in this
case the measure is positive definite at all stages of the
calculation, which is mathematically more rigorous.

Finally, we discuss an alternative way to regularize \eqref{eq:Z_2}
without diagonalizing $\D$:
\begin{align}
  Z(z,z^*;j) & =
  \big\langle{\det\big[(\D+z)(\D^\dagger+z^*)+j^2\big]}\big\rangle_{\ym}
  \label{eq:Z_2_j} \notag\\
  & =\left\langle{\det\begin{pmatrix}
      j & \D+z \\ \D^\dagger +z^* & -j
    \end{pmatrix}}\right\rangle_{\ym} \,.
\end{align}
The regulator $j$ has a clear physical meaning: It is a source for the
diquark condensate in two-color QCD (and for the pionic condensate in
QCD with nonzero isospin chemical potential, see
section~\ref{sec:iso}) \cite{Kanazawa:2011tt}.  Elementary
calculations lead to 
\begin{alignat}{1}
  &\bar{\del}\del \log Z(z,z^*;j) = \notag\\
  &\kakko{\tr\left[\frac{1}{(\D\!+\!z)(\D^\dagger\!+\!z^*)\!+\!j^2}j^2
      \frac{1}{(\D^\dagger\!+\!z^*)(\D\!+\!z)\!+\!j^2}\right]}_{\!N_f=2}
  \notag\\
  &+\bigg\langle\tr\left[(\D^\dagger+z^*)
    \frac{1}{(\D+z)(\D^\dagger+z^*)+j^2}\right] \notag\\
  &\quad\times \tr\left[(\D+z)\frac{1}{(\D^\dagger+z^*)(\D+z)+j^2}\right]
  \bigg\rangle_{N_f=2}\notag\\
  &-\kakko{\tr\left[(\D^\dagger\!+\!z^*)
      \frac{1}{(\D\!+\!z)(\D^\dagger\!+\!z^*)\!+\!j^2}\right]}_{N_f=2}
  \notag\\
  &\quad\times\kakko{\tr\left[(\D\!+\!z)
      \frac{1}{(\D^\dagger\!+\!z^*)(\D\!+\!z)\!+\!j^2}\right]}_{N_f=2}\,.
  \label{eq:j}
\end{alignat}
If the chiral limit $z\to0$ is taken at finite volume, the second and
third line vanish because of chiral symmetry.  Furthermore, it was
shown in appendix~A of \cite{Fyodorov1997} that at finite volume the
first term on the RHS becomes $\kakko{\pi\sum_n
  \delta^2(z+\lambda_n)}_{N_f=2}$ in the $j\to0$ limit.  However,
taking $j\to0$ before $V_4\to\infty$ does not give the desired result,
for the same reason that $\kappa\to0$ before $V_4\to\infty$ did not,
see the second comment above.  Unfortunately, we have been unable to
rewrite \eqref{eq:j} in terms of the eigenvalues of the Dirac operator
$\D$ due to its non-Hermitian nature and therefore cannot show
\eqref{eq:rho_main} explicitly.  This does not imply that the
correctness of the latter should be doubted.  We just cannot provide
an alternative derivation of \eqref{eq:rho_main} from \eqref{eq:j}.

\subsection{Consistency with the microscopic limit}
\label{sec:cons}

In analogy to \eqref{eq:one}, let us define the one-point function at
finite volume,
\begin{align}
  \bar\rho_1(\lambda) = \lim_{z\to 0}\frac1{V_4}
  \Big\langle{\sum_n \delta^2(\lambda-\lambda_n)}\Big\rangle_{N_f=2}\,.
\end{align}
In \cite{Kanazawa:2009ks} we proposed that the microscopic spectral
density defined by\footnote{Because of the change in conventions
  compared to \cite{Kanazawa:2009ks} (see footnote~\ref{foot:conv})
  there is now no factor of $1/V_4$ in front of $\rho_1$.}
\begin{align}
  \rho_s(\xi) & = \lim_{V_4\to\infty} \, \frac{\pi^2}{3\Delta^2}\,
  \bar\rho_1 \left(\frac{\pi}{\sqrt{3V_4\Delta^2}}\,\xi\right)
  \label{eq:rhos1}
\end{align}
is universal and can be computed in random matrix theory (RMT).  This
conjecture was consolidated in \cite{Kanazawa:2009en} by explicit
construction of the RMT corresponding to two-color QCD at high
density.  The function $\rho_s(\xi)$ itself was computed in
\cite{Akemann:2010tv} for arbitrary even $N_f$.  The explicit formula
for $\rho_s(\xi)$ in the limit of maximum non-Hermiticity reveals that
it asymptotically converges to an $N_f$-independent constant,
\begin{align}
  \lim_{|\xi|\to \infty} \rho_s(\xi) = \frac{1}{\pi}\,,
\end{align}
where we have to exclude purely real and purely imaginary values of
$\xi$.  This implies for large $|\xi|$
\begin{align}
  \frac{\pi^2}{3\Delta^2} \bar\rho_1
  \left(\frac{\pi}{\sqrt{3V_4\Delta^2}}\,\xi\right)
  \to \frac{1}{\pi}\,.
\end{align}
Note, however, that the argument of $\bar\rho_1$ goes to zero if
the limit $V_4\to\infty$ is taken before the limit
$|\xi|\to\infty$. Hence
\begin{align}
  \frac{\pi^2}{3\Delta^2} \rho_1(0)=\frac1\pi\,,
\end{align}
which exactly agrees with \eqref{eq:rho_main}.  Although this
``derivation'' is by no means rigorous, it is reassuring to see that
the microscopic limit is consistent with our main result.

\subsection{Consistency with spectral sum rules}
\label{sec:sumrules}

We now consider real quark masses.  If we are not interested in
deriving a Banks-Casher-type relation there is no reason to insist on
a positive measure, and we can take $m_1\ne m_2$.  We also omit the
regulator $\kappa$.

From \eqref{eq:logZ} we obtain with $z\to m_1$ and $z^*\to m_2$
\begin{align}
  \lim_{m_1,m_2\to0}\lim_{V_4\to\infty}
  \frac1{V_4}\frac{\del^2\log Z}{\del m_1\del m_2}
  =\frac{3}{\pi^2}\Delta^2\,.
  \label{eq:m1m2}
\end{align}
On the other hand, starting from the first line of \eqref{eq:Z_2}, and
using the fact that the Dirac eigenvalues come in pairs $\pm\lambda_n$
due to chiral symmetry, we have
\begin{align} 
  Z(m_1,m_2)
  & = \bigg\langle{\prod_n}'(-\lambda_n^2+m_1^2)
  (-\lambda_n^2+m_2^2)\bigg\rangle_{\ym} \,,
  \label{eq:Z2}
\end{align}
where the prime on the product (and on the sums below) means
$\re\lambda_n>0$.  This gives
\begin{align}
  &\frac{\del^2\log Z}{\del m_1\del m_2}
  =\bigg\langle{\sum_m}' \frac{2m_1}{-\lambda_m^2+m_1^2}
  {\sum_n}' \frac{2m_2}{-\lambda_n^2+m_2^2}\bigg\rangle_{N_f=2}\notag\\
  &\quad -
  \bigg\langle{\sum_n}'\frac{2m_1}{-\lambda_n^2+m_1^2}\bigg\rangle_{N_f=2}
  \bigg\langle{\sum_n}'\frac{2m_2}{-\lambda_n^2+m_2^2}\bigg\rangle_{N_f=2}\,.
  \label{eq:Zm1m2}
\end{align}
To compute the RHS we can make use of the massive spectral sum rules
derived in \cite{Kanazawa:2009ks} in the $\varepsilon$-regime of
chiral perturbation theory.  From eqs.~(4.32) and (4.33) of that
reference (with $\lambda\to i\lambda$ due to the change in
conventions), evaluated in the limit $m_1m_2V_4\Delta^2\to\infty$, we
find that the leading-order terms on the RHS of \eqref{eq:Zm1m2}
cancel and the next-to-leading-order term gives $3V_4\Delta^2/\pi^2$,
which is consistent with \eqref{eq:m1m2}.  This consistency check is
based on the fact that the $\varepsilon$- and the $p$-regime have
overlapping domains of validity \cite{Gasser:1987ah,Osborn:1998qb}.

Let us now consider $m_1=m_2=m$.  From \eqref{eq:logZ} we then obtain
\begin{align}
  \lim_{m\to0}\lim_{V_4\to\infty}\frac1{V_4}\,
  \frac{\del}{\del(m^2)}\log Z(m,m)
  =\frac{3}{\pi^2}\Delta^2+2\hh\,,
  \label{eq:mm}
\end{align}
while from \eqref{eq:Z2} with $m_1=m_2=m$ we obtain
\begin{align}
  \frac{\del}{\del(m^2)}\log Z(m,m)=
  \bigg\langle{\sum_n}'\frac2{-\lambda_n^2+m^2}\bigg\rangle_{N_f=2}\,.
  \label{eq:bad_sumrule}
\end{align}
To evaluate the RHS we use \cite[eq.~(4.32)]{Kanazawa:2009ks},
evaluated in the limit $m^2V_4\Delta^2\to\infty$, and obtain
$3V_4\Delta^2/\pi^2$.  This is not consistent with \eqref{eq:mm} since
the $\hh$ term is missing.  The reason for this discrepancy is that in
\cite{Kanazawa:2009ks} the high-energy term $\hh\tr(M^2)$ was not
considered.  Including this term modifies some of the spectral sum
rules and makes \eqref{eq:bad_sumrule} consistent with \eqref{eq:mm}.
This issue is discussed in more detail in appendix~\ref{app:sumrules}.

\subsection[General $N_f$]{\boldmath General $N_f$}
\label{sec:general}

So far we have worked with two flavors, and we now show that our main
result is unchanged for general even $N_f$.  We have nothing to say
about odd $N_f$ since the measure is not positive definite in this
case.  The mass matrix is now $M=\diag(z,\ldots,z,z^*,\ldots,z^*)$
with $N_f/2$ entries each of $z$ and $z^*$.  The generalization of
\eqref{eq:dE2} to even $N_f\ge4$ is given by
\cite{Kanazawa:2009ks}\footnote{See the discussion before
  \eqref{eq:dE2}.  In \cite{Kanazawa:2009ks} we assumed $M_{LR}=M$ and
  $M_{RL}=M^\dagger$, while we now have $M_{LR}=M_{RL}=M$.  This
  implies that additional terms in the chiral Lagrangian are allowed
  by symmetries, e.g., $A^2\tr(M\Sigma_RM^*\Sigma_L^\dagger)$.
  However, HDET shows that the coefficients of these terms vanish and
  that only the terms in \eqref{eq:dE_Nf} remain.  See also
  \cite{Son:1999cm,Son:2000tu} for another example where terms allowed
  by symmetries vanish in HDET.  Note also that for the values of
  $\Sigma_R$ and $\Sigma_L$ that maximize \eqref{eq:dE_Nf} (see below)
  the term in $\{\cdots\}$ is already
  real.\label{foot:other}}\footnote{Similarly to footnotes
  \ref{foot:het} and \ref{foot:problem} for $N_f=2$, the axial anomaly
  (or instantons) gives rise to another mass term at $\mO(m^2)$ in the
  free energy for $N_f\ge4$, which is not displayed in
  \eqref{eq:dE_Nf} because it vanishes at sufficiently high
  density. But it is nonzero at lower density.  This mass term is
  obtained from the instanton-induced interaction with $N_f$
  left-handed and $N_f$ right-handed fermion legs, by replacing two
  sets of left- and right-handed fermions by $M^2$ and the remaining
  $2(N_f-2)$ fermions by diquark condensates, in a way analogous to
  \cite{Schafer:2002ty,Yamamoto:2008zw}.}
\begin{align}
  \label{eq:dE_Nf}
  \delta\mE=-\frac{3\Delta^2}{4\pi^2}\max_{A,\Sigma_L,\Sigma_R}
  & \re \Big\{
    A^2\tr(M\Sigma_RM^T\Sigma_L^\dagger) \notag\\
    & \quad\:\, + A^{*2}\tr(M\Sigma_LM^T\Sigma_R^\dagger)
  \Big\}\,,
\end{align}
where $A\in\U(1)$, $\Sigma_L\in\SU(N_f)_L/\Sp(N_f)_L$ and
$(L\leftrightarrow R)$.  An explicit parameterization of
$\Sigma_{L/R}$ is given in \cite{Kanazawa:2009ks}, but it is
unimportant for our present purpose.  Note that the minimum has become
a maximum together with an overall minus sign.  We now use a
Cauchy-Schwarz inequality for matrices,
\begin{align}
  |\tr(X^\dagger Y)|^2\le\tr(X^\dagger X)\tr(Y^\dagger Y)\,,
\end{align}
which becomes an equality if and only if $X$ and $Y$ are multiples of
each other \cite{abadir2005matrix}.  Because of
$\Sigma_L^\dagger\Sigma_L=\Sigma_R^\dagger\Sigma_R=\1$ this leads to
\begin{align}
  \label{eq:CS}
  |\tr(M\Sigma_RM^T\Sigma_L^\dagger)|\le\tr(M^\dagger M)=N_fzz^*\,,
\end{align}
where for our choice of $M$ the maximum is obtained\footnote{Note that
  the maximum could also be obtained for
  {$\Sigma_L=\Sigma_R=\pm\begin{pmatrix}0&\1\\\1&0\end{pmatrix}$}, but
  such $\Sigma$ are not elements of the coset space
  $\SU(N_f)/\Sp(N_f)$ and therefore excluded.} for
$\Sigma_L=\Sigma_R=\pm I$ with
\begin{align}
  \label{eq:I}
  I=\begin{pmatrix}
    0 & -\1_{N_f/2}\\
    \1_{N_f/2} & 0
  \end{pmatrix}.
\end{align}
Together with $A=\pm1$ we thus have
\begin{align}
  \delta\mE=-\frac{3N_f}{2\pi^2}\Delta^2 zz^*\,.
\end{align}
Including the high-energy term $\hh\tr(M^2)$ we obtain
\begin{align}
  \frac1{V_4}\log Z(z,z^*)
  =\frac{N_f}2\bigg[\frac{3\Delta^2}{\pi^2}zz^*
  \!+\! \hh(z^2+z^{*2})\bigg]+\mO(|z|^3)\,.
  \label{eq:ZNf}
\end{align}
Thus the only difference to \eqref{eq:logZ} is a factor of $N_f/2$,
and the analog of \eqref{eq:LHS} is
\begin{align}
  \label{eq:LHS_Nf}
  \lim_{z\to0}\lim_{V_4\to\infty}\frac1{V_4}\,
  \bar\del\del\log Z(z,z^*)=\frac{3N_f}{2\pi^2}\Delta^2\,.
\end{align}
The analog of \eqref{eq:Z_2_kappa} is
\begin{align}
  Z(z,z^*;\kappa) =
  \Big\langle\prod_n\big[(\lambda_n+z)(\lambda_n^*+z^*)
    +\kappa^2 \big]^{N_f/2}\Big\rangle_{\ym}\,,
\end{align}
and going through the same steps as in section~\ref{sec:deriv} we
obtain 
\begin{align}
  \label{eq:QCD_Nf}
  \lim_{\kappa\to 0} \lim_{z\to 0} \lim_{V_4\to \infty}
  \frac{1}{V_4}\,\bar{\del}\del \log Z(z,z^*;\kappa) 
  = \frac{N_f}2\,\pi {\rho}_1^{(N_f)}(0)
\end{align}
with the one-point function in the presence of $N_f$ massless flavors.
Hence the Banks-Casher-type relation for $N_f\ge4$ is the same as for
two flavors.  The arguments of sections~\ref{sec:cons} to
\ref{sec:sumrules} also generalize straightforwardly to general even
$N_f$.

\section{QCD at nonzero isospin density}
\label{sec:iso}

Let us now consider QCD with two flavors and any number of colors
$N_c\ge2$ at large isospin chemical potential $\mu_I=2\mu$.  In this
case we have
\begin{align}
  Z(m_1,m_2)
  & = \kakko{  \det(\D(\mu)+m_1)\det(\D(-\mu)+m_2) }_{\ym}
  \notag\\
  & = \big\langle{\det(\D+m_1)\det(\D^\dagger+m_2)}\big\rangle_{\ym}\,,
  \label{eq:Z_iso} 
\end{align}
where the last line follows from $D(\mu)^\dagger=-D(-\mu)$ and the
fact that the eigenvalues of $D$ come in pairs $\pm\lambda$.  This is
the same as \eqref{eq:Z_2}.  Setting $M=\diag(z,z^*)$ we again start
from \eqref{eq:Z_2_kappa} and go through the same steps as in
section~\ref{sec:deriv} to obtain \eqref{eq:Z_rho_main}.

In QCD at sufficiently large $\mu_I$, BCS pairing of the form $\langle
\bar d \gamma_5 u \rangle$ is formed near the Fermi surface according
to the BCS mechanism \cite{Son:2000xc,Son:2000by}.  The chiral
Lagrangian in this regime can be constructed similarly to
\cite{Kanazawa:2009ks}, but the coset space is now
$\U(1)_A\times\U(1)_{I_3}$ \cite{Kanazawa:PhD}, where the latter is
the $\U(1)$ symmetry with respect to the third isospin generator.
Repeating the analysis in \cite{Kanazawa:2009ks} with the modification
discussed before \eqref{eq:dE2}, the mass term can be written as
\cite[eq.~(4.13)]{Kanazawa:PhD}
\begin{align}
  \mL_\text{mass}=-c_\text{iso} \Delta^2 
  \left\{(A^2 + A^{*2})\det M
  \right\},
\end{align}
where $A\in\U(1)$ and $c_\text{iso}$ is a positive coefficient, see
\eqref{eq:ciso} below.  Thus
\begin{align}
  \delta\mE 
  =-c_\text{iso} \Delta^2\max_{A}
  \left\{(A^2 + A^{*2})\det M \right\}
  = - 2 c_\text{iso} \Delta^2 zz^*\,,
  \label{eq:Eiso}
\end{align}
where the last equation is obtained with $A=\pm 1$.

To compute the coefficient $c_\text{iso}$ we use essentially the same
arguments as for the color-flavor-locked (CFL) phase
\cite{Schafer:2001za} and for dense two-color QCD
\cite{Kanazawa:2009ks}, the only new feature being that the sign of
the chemical potential is opposite for the two flavors.  A simple way
to adapt QCD with isospin chemical potential to the method of
\cite{Schafer:2001za,Kanazawa:2009ks} is to change the basis of the
quark fields from $(u,d)$ to $(u,d_c)$ with $d_c \equiv C\bar{d}{}^T$.
Accordingly, the Lagrangian is then rewritten as
\begin{align}
  \mL &= \bar{u}[D(\mu)+m_1]u+\bar{d}[D(-\mu)+m_2]d \notag\\
  & = \bar{u}[D(\mu)+m_1]u+
  \bar{d}_c[D_\text{AF}(\mu)+m_2]d_c\,,
\end{align}
where $D_\text{AF}$ is the Dirac operator in the anti-fundamental
representation of $\SU(N_c)$.  Adapting the mass term in HDET, see,
e.g., \cite[eq.~(A.1)]{Kanazawa:2009ks}, to the present case we
obtain\footnote{The notation in \eqref{eq:L_iso} is a bit sloppy since
  $C$ is a $4\times4$ matrix, while the right- and left-handed spinors
  are two-spinors, which here are considered to be four-spinors with
  two components set to zero.  Choosing $C=i\gamma_4\gamma_2$ we have
  $(d_c^\dag)_{L}^{a}C {u_{L}^*}^b\to (d_c^\dag)_{L}^{a}\sigma_2
  {u_{L}^*}^b=(d_R^{\dag a} u_L^b)^*$ and $(d_c^T)_{R}^c C u_{R}^d\to
  -(d_c^T)_{R}^c \sigma_2 u_{R}^d =d_L^{\dag c}u_R^d$.}
\begin{align}
  \label{eq:L_iso}
  \mL_\text{mass}^\text{HDET} &= \frac{g^2}{8\mu^4}
    \Big((\psi_L^\dag)_{i}^{a}C (\psi_L^*)_{j}^b \Big)
    \Big((\psi_R^T)_{k}^{c} C (\psi_R)_{l}^d \Big) \notag\\
    &\quad \times(\tilde T^A)^{ac} (T^A)^{bd}M_{ik}M_{jl} 
    + (L\leftrightarrow R)\,,
\end{align}
where $\psi_1=u$ and $\psi_2=d_c$, $g$ is the coupling constant,
$a,b,c,d$ are color indices, $i,j,k,l$ are flavor indices, and the
$\tilde T^A\equiv -(T^A)^*$ are the generators of $\SU(N_c)$ in the
anti-fundamental representation with normalization
$\tr(T^AT^B)=\delta^{AB}/2$.  To leading order in $g^2$, the
expectation value of \eqref{eq:L_iso} factorizes into the product of
the expectation values of the diquark condensates, and the latter can
be computed in a weak-coupling calculation \cite{Schafer:1999fe}.  At
high isospin density, BCS-type diquark pairing occurs in the
color-symmetric, flavor-antisymmetric, and pseudoscalar channel, i.e.,
the quantum numbers of the condensate are the same as for pions
\cite{Son:2000xc,Son:2000by}.  We therefore have
\begin{align}
  \label{eq:pcond}
  \kakko{(\psi_L^T)_{i}^{a} C (\psi_L)_{j}^b}_{N_f=2}
  =\delta^{ab}\eps_{ij}r_LA_L
  \quad\text{and}\quad(L\leftrightarrow R)\,,
\end{align}
where $r_{L/R}\in\mathds{R}_{\ge0}$ and $A_{L/R}\in\U(1)$ are
magnitude and phase of the pionic condensate.  For the magnitude we
can use the result \cite[eq.~(3.42)]{Hanada:2011ju}, which in the
present context reads
\begin{align}
  \big|\kakko{d_c^TC\gamma_5u}_{N_f=2}\big|
  &= 2\sqrt{\frac{6N_c}{N_c^2-1}}\frac{\mu^2\Delta}{\pi g} \,,\notag\\
  \big|\kakko{d_c^TCu}_{N_f=2}\big|&=0\,.
\end{align}
This implies
\begin{align}
  \label{eq:hy}
  r_L=r_R=\sqrt{\frac{6N_c}{N_c^2-1}}\frac{\mu^2\Delta}{\pi g}\,.
\end{align}
Taking the expectation value of \eqref{eq:L_iso} and noting that
$\delta^{ab}\delta^{cd} (\tilde T^A)^{ac}
(T^A)^{bd}=-\tfrac12(N_c^2-1)$ we thus obtain with $U=A_L^\dag A_R$
\begin{align}
  \delta\mE & = -\frac{3N_c}{4\pi^2}\Delta^2 \max_U\big\{(U+U^*)\det
  M\big\} = -\frac{3N_c}{2\pi^2}\Delta^2zz^*
\end{align}
with $U=1$ in the last step.  Comparing with \eqref{eq:Eiso} we find
\begin{align}
  \label{eq:ciso}
  c_\text{iso}=\frac{3N_c}{4\pi^2}\,.
\end{align}
Including the high-energy term $\hh\tr(M^2)$ we thus have
\begin{align}
  \label{eq:logZ_iso}
  \frac{1}{V_4}\log Z(z,z^*)= 
  \frac{3N_c}{2\pi^2} \Delta^2 zz^*
  + \hh (z^2+z^{*2})+\mO(|z|^3)\,.
\end{align}
From \eqref{eq:Z_rho_main} and \eqref{eq:logZ_iso} we obtain the
Banks-Casher-type relation
\begin{align}
  \label{eq:rho_main2}
  \Delta^2 = \frac{2\pi^3}{3N_c}\rho_1(0)\,. 
\end{align}
For $N_c=2$ \eqref{eq:rho_main2} reduces to \eqref{eq:rho_main} as it
should (recall that there is no difference between isospin chemical
potential and baryon chemical potential for $N_c=2$
\cite{Splittorff:2000mm}).

In analogy with section~\ref{sec:cons} we can check the consistency of
\eqref{eq:rho_main2} with the microscopic limit.  For this consistency
check we need to map the low-energy effective theory to RMT.  This
mapping was derived in \cite[Sec.~4.2]{Kanazawa:PhD} with the
exception of a numerical prefactor, which can be fixed by matching
\eqref{eq:logZ_iso} and \cite[Eq.~(4.37)]{Kanazawa:PhD}.  Using this
prefactor in \cite[Eq.~(4.49)]{Kanazawa:PhD} results in the definition
\begin{align}
  \rho_s(\xi) & = \lim_{V_4\to\infty} \frac{2\pi^2}{3N_c\Delta^2}\,
  \bar\rho_1\Bigg( \sqrt{\frac{2\pi^2}{3N_cV_4\Delta^2}} \ \xi \Bigg).
  \label{eq:rhos2}
\end{align}
Adapting the analytical result for $\rho_s(\xi)$
\cite[Eq.~(4.51)]{Kanazawa:PhD} with rescaled quark masses $\tau_1$\
and $\tau_2$ to our case gives
\begin{align}
  \rho_s(\xi) & = \frac{2}{\pi}|\xi|^2 K_0(|\xi|^2) \left(
    I_0(|\xi|^2) - \frac{I_0(\tau_1\xi^*)I_0(\tau_2\xi)}{I_0(\tau_1\tau_2)}
  \right)
  \notag\\
  & \to \frac{1}{\pi}\quad \text{as }|\xi|\to \infty \,,
  \label{eq:cons2}
\end{align}
which is consistent with \eqref{eq:rho_main2}.

\section{Adjoint QCD}
\label{sec:adj}

We now derive the Banks-Casher-type relation for QCD with fermions in
the adjoint representation.  To obtain the Dirac spectrum in adjoint
QCD one might naively want to consider the $N_f=1$ partition function
$\kakko{\det(D+z)}$, with $D$ in the adjoint representation of
$\SU(N_c)$, but for our purposes this is not a sensible choice because
the positivity of the measure is spoiled for complex $z$.\footnote{For
  real mass mass $z=m$ the measure is actually positive in $N_f=1$
  adjoint QCD so that in principle one could derive a
  Banks-Casher-type relation.  However, in this case the terms
  involving $\Delta^2$ and $H_2$ have the same mass dependence so that
  $\Delta^2$ alone cannot be extracted by taking derivatives with
  respect to $m$.}  For this reason we consider an even number $N_f$
of flavors for which we can insert complex conjugate masses, i.e., the
mass matrix is $M=\diag(z,\ldots,z,z^*,\ldots,z^*)$ as in
section~\ref{sec:general} and
\begin{align}
  Z(z,z^*) 
  &= \kakko{{\det}^{N_f/2}(D+z)\,{\det}^{N_f/2}(D+z^*)}_{\ym}\notag\\
  &= \kakko{{\det}^{N_f/2}(D+z)\,{\det}^{N_f/2}(D^\dag+z^*)}_{\ym}\,,
\end{align}
where the last equality follows from the anti-unitary symmetry
$[C\gamma_5K,D]=0$ of the adjoint Dirac operator
\cite{Leutwyler:1992yt}, which implies
$D^*=(C\gamma^5)^{-1}D(C\gamma_5)$.  On the QCD side we are then led
to \eqref{eq:QCD_Nf} as described before for two-color QCD.

We now turn to the HDET side.  The construction of the chiral
Lagrangian proceeds in close analogy to \cite{Kanazawa:2009ks}, the
difference being that the coset space is now
$\U(1)_B\times\U(1)_A\times[\SU(N_f)_L/\SO(N_f)_L]\times
[\SU(N_f)_R/\SO(N_f)_R]$ \cite{Kanazawa:2011tt}.  Repeating the
analysis in \cite{Kanazawa:2009ks} with the modification discussed
before \eqref{eq:dE2} we obtain \cite{Kanazawa:PhD}\footnote{As in
  footnote~\ref{foot:other} additional terms are allowed by
  symmetries, but their coefficients are zero in HDET.}
\begin{align}
  \mL_\text{mass}=-c_\text{adj} \Delta^2 
  \Big\{ &
  A^2\tr( M\Sigma_RM^T\Sigma_L^\dag) \notag\\
  & + A^{*2}\tr(M\Sigma_LM^T\Sigma_R^\dagger)
  \Big\}\,,
\end{align}
where $A\in\U(1)$, $\Sigma_L\in\SU(N_f)_L/\SO(N_f)_L$ and
$(L\leftrightarrow R)$, and $c_\text{adj}$ is a positive coefficient,
see \eqref{eq:cadj} below.  Thus
\begin{alignat}{2}
  \delta\mE 
  & =-c_\text{adj} \Delta^2\max_{A,\Sigma_L,\Sigma_R}
  \re \Big\{ && A^2\tr( M\Sigma_RM^T\Sigma_L^\dag) \notag\\
  &&& + A^{*2}\tr(M\Sigma_LM^T\Sigma_R^\dagger) \Big\}
  \notag \\
  \label{eq:Eadj}
  & = -c_\text{adj} \Delta^2\cdot2N_fzz^*\,,
\end{alignat}
where we have employed the argument leading to
\eqref{eq:CS}.\footnote{The maximum is now obtained for $A=\pm1$ and
  $\Sigma_L=\Sigma_R=\pm
  \begin{pmatrix}0&\1_{N_f/2}\\\1_{N_f/2}&0\end{pmatrix}$.}  

The coefficient $c_\text{adj}$ can again be computed from HDET.
Instead of \eqref{eq:L_iso} we now have
\begin{align}
  \mL_\text{mass}^\text{HDET} &= \frac{g^2}{8\mu^4}\Big( 
    (\psi^\dag_L)_{i}^{a}C(\psi^*_L)_{j}^{b}\Big)
    \Big((\psi_R^T)_{k}^{c} C (\psi_R)_{l}^d\Big) \notag\\
    &\quad\times 
    (t^A)^{ac} (t^A)^{bd} M_{ik}M_{jl} + (L\leftrightarrow R) \,,
    \label{eq:Eshift}
\end{align}
where the $t^A$ $(A=1,\dots,N_c^2-1)$ are the generators of $\SU(N_c)$
in the adjoint representation, i.e., $(t^a)^{bc}=-if^{abc}$ with the
structure constants $f^{abc}$.  To leading order in $g^2$ the
expectation value of \eqref{eq:Eshift} again factorizes into the
product of the expectation values of the diquark condensates.  The
condensation again occurs in the pseudoscalar channel, but another
difference to \cite{Kanazawa:2009ks} is that the diquark condensate is
now symmetric in the color and flavor indices (and antisymmetric in
the spin indices as before), i.e.,
\begin{align}
  \kakko{(\psi_L^T)_i^a C (\psi_L)_j^b}_{N_f}
  =\delta^{ab}\delta_{ij}r_{L}A_{Li}
  \quad \text{and}\quad (L\leftrightarrow R)\,,
\end{align}
where $r_{L/R}\in\mathds{R}_{\ge0}$ and $A_{(L/R)i}\in\U(1)$ are the
magnitudes and phases of the condensates.  From weak-coupling
calculations at high density we find\footnote{This can be obtained by
  replacing the group theoretical factor in the gap equation in
  \cite{Schafer:1999fe,Hanada:2011ju} by $(f^{a})^{bc} ({\1})^{cd}
  (f^a)^{de} = -N_c({\1})^{be}$.}
\begin{equation}
  \label{eq:acond}
  r_L=r_R=\sqrt{\frac{3}{N_c}} \frac{\mu^2\Delta}{\pi g} \,.
\end{equation}
While the magnitude of the condensates is fixed regardless of the mass
term, the phases of the condensates must be such that the ground state
energy is minimized.  Taking the expectation value of
\eqref{eq:Eshift} and using $\delta^{ab}\delta^{cd}(t^A)^{ac}
(t^A)^{bd}=-N_c(N_c^2-1)$ we thus obtain
\begin{align}
  \label{eq:Eadj2}
  \delta\mE&=- \frac{3(N_c^2-1)}{8\pi^2} \Delta^2 \notag\\
  & \quad \times \max_{A_L,A_R} \re \Big\{ \tr(A_L^\dagger MA_RM^T) 
  + (R\leftrightarrow L)\Big\}\notag\\
  &=- \frac{3(N_c^2-1)}{8\pi^2} \Delta^2\cdot 2N_fzz^*\,,
\end{align}
where $A_L=\diag(A_{L1},\ldots,A_{LN_f})$ and $(L\leftrightarrow R)$,
and in the last step we have once again used the argument leading to
\eqref{eq:CS}.  Note that the maximum is now obtained for
$A_R=M^{-1}A_LM^*$.\footnote{This assumes that $M$ is diagonal since
  for general $M$ the combination $M^{-1} A_L M^*$ would not be
  diagonal.}  Comparing with \eqref{eq:Eadj} we find
\begin{align}
  \label{eq:cadj}
  c_\text{adj}=\frac{3(N_c^2-1)}{8\pi^2}\,.
\end{align}
Including again the high-energy term $H_2 \tr(M^2)$ we have
\begin{align}
  &\frac{1}{V_4}\log Z(z,z^*) = \notag\\
  &\quad \frac{3(N_c^2-1)}{4\pi^2}N_f \Delta^2zz^* 
  + \hh (z^2+z^{*2})+\mO(|z|^3)\,. 
  \label{eq:Zadj}
\end{align}
Comparing with \eqref{eq:QCD_Nf} we arrive at the Banks-Casher-type
relation for adjoint QCD,
\begin{align}
  \Delta^2 = \frac{2\pi^3}{3(N_c^2-1)}\rho_1(0)\,.
  \label{eq:BC_adj}
\end{align}

Let us add an important remark here.  In deriving the RHS of
\eqref{eq:Eadj2} it was essential to include the phases of the
condensates.  The minimum of $\delta\mE$ is obtained only if some of
the phases are nontrivial, i.e., $\delta\mE$ would not have been
minimized if all phases had been set to unity.  In contrast, in the
other two cases (two-color QCD and isospin) the minimum of $\delta\mE$
is obtained for trivial phases, but this should better be regarded as
accidental.

We again check the consistency of \eqref{eq:BC_adj} with the
microscopic limit.  The mapping between the low-energy effective
theory of dense adjoint QCD and RMT was derived in
\cite[Sec.~4.3]{Kanazawa:PhD}.  The corresponding microscopic spectral
density $\rho_s(\xi)$ is a special case (i.e., the limit of maximum
non-Hermiticity) of a more general result computed by Akemann
\cite{Akemann:2005fd}.  The proper definition of $\rho_s(\xi)$ is
obtained by matching \eqref{eq:Zadj} for $N_f=2$ with
\cite[Eq.~(4.17)]{Akemann:2005fd} and using the resulting factor in
\cite[Eq.~(4.9)]{Akemann:2005fd}.  This gives\footnote{Here and in
  \eqref{eq:rho_s_Ake}, our $\rho_s(\xi)$ differs from that of
  \cite{Akemann:2005fd} by a factor of 4, which is due to a
  different normalization of $\rho_1(\lambda)$ in
  \cite{Akemann:2005fd} and here.}
\begin{align}
  \rho_s(\xi) & = \lim_{V_4\to\infty}
  \frac{2\pi^2}{3(N_c^2-1)\Delta^2}\,\bar\rho_1 \left(
    \sqrt{\frac{2\pi^2}{3(N_c^2-1)V_4\Delta^2}}\ \xi
  \right).
  \label{eq:rhos4}
\end{align}
Since the number of flavors does not affect the asymptotic behavior of
$\rho_s(\xi)$ it is sufficient for our purpose to use the quenched
($N_f=0$) result at maximum non-Hermiticity
\cite[Eq.~(4.10)]{Akemann:2005fd},
\begin{align}
  \rho_s(\xi) & = \frac{1}{\pi}(\xi^{*2}-\xi^2)|\xi|^2 K_{0}(|\xi|^2) 
  \notag 
  \\
  & \quad \times 
  \int_0^1 \!\frac{dr}{\sqrt{1-r^2}}I_{0}(r|\xi|^2)
  \sinh\Big( \frac{1}{2}\sqrt{1-r^2}(\xi^2\!-\!\xi^{*2}) \Big)
  \label{eq:rho_s_Ake}
  \notag\\
  & \to \frac{1}{\pi}\quad \text{as }|\xi|\to\infty\,,
\end{align}
which is consistent with \eqref{eq:BC_adj}.  Note that the definitions
\eqref{eq:rhos1}, \eqref{eq:rhos2}, and \eqref{eq:rhos4} can be
summarized as
\begin{align}
  \rho_s(\xi) & = \lim_{V_4\to\infty}
  \frac{2\pi^2}{3d_\text{rep}\Delta^2}\,\bar\rho_1 \left(
    \sqrt{\frac{2\pi^2}{3d_\text{rep}V_4\Delta^2}}\ \xi \right).
\end{align}

\section{Conclusion and outlook}
\label{sec:summary}
In this paper, we have derived a new Banks-Casher-type relation at
high density and zero temperature that relates the complex Dirac
eigenvalue density at the origin to the BCS gap squared. We have shown
that the proportionality constant in this relation is independent of
the number of flavors and is fixed by weak-coupling calculations.
(The latter point should be contrasted with the usual Banks-Casher
relation in the QCD vacuum, where the proportionality constant is
fixed by the definition of the chiral condensate.)  We have also
checked the consistency of our relation with previous results for the
spectral density of Dirac eigenvalues in the microscopic limit
\cite{Kanazawa:2009ks,Akemann:2010tv,Kanazawa:PhD,Akemann:2005fd}.

As mentioned in the introduction, the positivity of the measure is a
necessary condition to derive the Banks-Casher-type relation.  This is
why its generalization to the CFL color superconducting state
\cite{Alford:1998mk} in QCD at high baryon density is not feasible,
though the Dirac eigenvalue spectrum itself is related to the BCS gap
\cite{Yamamoto:2009ey}.  Still, it is likely that the BCS gap of the
CFL phase is characterized by some singular behavior of the Dirac
spectrum near the origin, like the scenario in
\cite{Osborn:2005ss,Osborn:2008jp}.  This would be an interesting
question to explore.  One should also be able to generalize our
Banks-Casher-type relation to nonzero temperature by solving the gap
equation and computing the quark-mass dependence of the ground-state
energy in terms of the BCS gap at nonzero temperature.

Our relation enables us to compute the magnitude of the BCS gap
numerically through the low-lying Dirac eigenvalue spectrum on the
lattice.  More detailed information on the Dirac spectrum is provided
by the microscopic spectral density (and higher-order spectral
correlations) of the Dirac eigenvalues at high density.  These
quantities have been computed for two-color QCD \cite{Akemann:2010tv},
and it would be interesting to compute them for the isospin and
adjoint cases as well.  Combined with the measurement of the diquark
(or pionic) condensate that can be obtained from the Dirac
\emph{singular value} spectrum \cite{Kanazawa:2011tt}, one should be
able to perform a detailed quantitative analysis at high density
numerically.  We hope that such calculations from first principles
will eventually lead to solid evidence that the BCS mechanism, and
moreover the BEC-BCS crossover phenomena, ubiquitously occur not only
in condensed matter systems but also in high-energy physics.

\begin{acknowledgement}
  TK was supported by the Alexander von Humboldt foundation.  TW is
  supported by DFG (SFB/TRR-55).  NY was supported in part by JSPS
  Postdoctoral Fellowships for Research Abroad and JSPS Research
  Fellowships for Young Scientists.
\end{acknowledgement}

\appendix

\section{Computation of the high-energy constant \boldmath$\hh$}
\label{app:h2}

The term $\hh\tr(M^2)$ in \eqref{eq:logZ} and \eqref{eq:ZNf} was
originally introduced in chiral perturbation theory as a counterterm
needed in the renormalization of one-loop graphs
\cite{Gasser:1983yg,Gasser:1984gg}.  For the same reason it is
necessary in the chiral perturbation theory of two-color QCD at high
density.  At sufficiently high density where $g \ll 1$, the
coefficient $\hh$ can be determined by computing the ground-state
energy density microscopically.  In this appendix, we compute $\hh$ to
leading order in $g$ (i.e., in the free theory).  For this purpose, we
use the massless quark propagator with chemical potential
$\mu$,\footnote{Note that the contour in the usual Feynman
  prescription is modified from $\mu=0$ to $\mu \neq 0$
  \cite[fig.~4]{Shuryak:1980tp}.  For $\mu=0$, \eqref{eq:prop} reduces
  to the conventional $i\epsilon$ prescription because $p_0\,
  \sgn(p_0) > 0$.}
\begin{align}
  \label{eq:prop}
  &\frac{1}{\slashed{p} + \mu \gamma_0}
  =
  \frac{(p_0 + \mu ) \gamma_0 - {\bm p} \cdot {\bm \gamma}}
  {[p_0 + \mu  + i \epsilon \sgn (p_0) ]^2 - {\bm p}^2}
  \notag\\
  &\quad=
  \frac{(p_0+\mu)\gamma_0 - {\bm p} \cdot {\bm \gamma}}
  {p_0 + \mu + p  - i \epsilon} \notag\\
  &\qquad\times\bigg[\frac{\theta(p-\mu)}{p_0 - (p-\mu) + i \epsilon}
  + \frac{\theta(\mu-p)}{p_0 - (p-\mu) - i \epsilon} \bigg], 
\end{align}
where sgn$(x)$ denotes the sign of $x$ and $p=|{\bm p}|$.  A one-loop
calculation gives the free energy due to the (degenerate) quark
masses,
\begin{equation}
  \frac{Z(m)}{Z(0)} = 1 + \frac{(-1)}{2} 
  V_4 m^2 \int \frac{d^4 p}{(2\pi)^4}\,\Tr
  \left(\frac{1}{\slashed{p} + \mu \gamma_0}\right)^2,
\end{equation}
where the factor $(-1)$ is due to a fermion loop and ``$\Tr$" is the
trace over spinor, color, and flavor indices.  When the integration
contour is closed in the upper half-plane, the integrand has two poles
at $p_0 = \pm p - \mu$.  From \eqref{eq:prop}, the pole at $p_0 = p -
\mu$ gives rise to the $\mu$-dependent contribution, while $p_0 = - p
- \mu$ gives the $\mu$-independent vacuum contribution.  Performing
the integration over $p$, the $\mu$-dependent part reads
\begin{align}
  & \mE_{\mu \neq 0}(m) \notag\\
  &= -\frac{1}{2} V_4 m^2
  \int_0^{\mu} \frac{d^3 p}{(2\pi)^3}\tr\left[\frac{d}{d p_0}
    \frac{4[(p_0+\mu)^2 - p^2]}{(p_0 + p + \mu)^2} \right]_{p_0=p-\mu}
  \notag\\
  &= - d_\text{rep} N_f V_4 \frac{\mu^2 m^2}{4\pi^2}\,,
\end{align}
where ``$\tr$'' is now the trace over color and flavor indices and
$d_\text{rep}$ is the dimension of the fermion representation.  On the
other hand, the vacuum contribution is given by
\begin{align}
  \label{eq:Z_pt}
  \mE_{\mu=0}(m) &= -\frac{1}{2}V_4m^2\int\frac{d^4p}{(2\pi)^4}\,\Tr
  \frac{1}{\slashed{p}^2} \notag\\
  &=  - d_\text{rep} N_f V_4 \frac{\Lambda_\text{UV}^2m^2}{4\pi^2}\,,
\end{align}
where we have regularized the divergent integral by introducing a
momentum cutoff $\Lambda_\text{UV}$.  Note that this term depends on
the scheme used to define the finite part of the diagrams.  For
instance, dimensional regularization or Pauli-Villars regularization
would lead to different values.  If we had used dimensional
regularization the divergent integral in \eqref{eq:Z_pt} would simply
be zero.

Putting the terms together and comparing with \eqref{eq:ZNf}, we
obtain
\begin{equation}
  \hh = - \frac{d_\text{rep}}{4\pi^2} \left(\mu^2+\Lambda_\text{UV}^2\right),
\end{equation}
where the second term depends on the scheme we employ.  This result is
also valid for the other cases (QCD with isospin chemical potential
and adjoint QCD).

\section{\boldmath Modified spectral sum rules}
\label{app:sumrules}

Spectral sum rules for the inverse Dirac eigenvalues were first
derived by Leutwyler and Smilga \cite{Leutwyler:1992yt} for QCD with
massless fermions at zero density.  These massless sum rules were then
generalized to theories with real and pseudoreal fermions by Smilga
and Verbaarschot \cite{Smilga:1994tb}.  The extension of all three
cases to massive fermions was given by Damgaard and Splittorff
\cite{Damgaard:1997cy,Damgaard:1999ic}.  In \cite{Kanazawa:2009ks} we
derived massless and massive spectral sum rules for two-color QCD
(i.e., the case of pseudoreal fermions) at high density.  The purpose
of this appendix is to discuss modifications of these spectral sum
rules due to the high-energy constant $\hh$ that was not taken into
account in \cite{Kanazawa:2009ks}.  For simplicity we focus on
two-color QCD at high density with $N_f=2$ and real masses. In the
$\epsilon$-regime at high density we have from \eqref{eq:logZ} and
\cite{Kanazawa:2009ks}
\begin{align}
  \label{eq:Zapp}
  Z(m_1,m_2) = I_0\Big(\frac{3}{\pi^2}V_4\Delta^2m_1m_2\Big) 
  \exp\big[V_4\hh(m_1^2+m_2^2)\big]\,,
\end{align}
where $I_0$ is a modified Bessel function and the second factor is
missing in \cite{Kanazawa:2009ks}.  This is the most generic form of
the partition function at leading order of the $\epsilon$-expansion,
consistent with all symmetries of the system.

We now investigate how the inclusion of the high-energy term modifies
the spectral sum rules derived in \cite{Kanazawa:2009ks}.  Taylor
expansion of \eqref{eq:Zapp} gives
\begin{align}
  & Z(m_1,m_2) = 1 + V_4 \hh (m_1^2+m_2^2) 
  + \alpha(V_4\Delta^2)^2m_1^2m_2^2 \notag\\
  & + \frac{1}{2}(V_4\hh)^2 (m_1^2+m_2^2)^2
  + \frac{1}{6}(V_4\hh)^3 (m_1^2+m_2^2)^3 \notag\\
  & + \alpha(V_4\Delta^2)^2(V_4\hh)m_1^2m_2^2(m_1^2+m_2^2) 
  + \mO(m^8)
  \label{eq:Zeff}
\end{align}
with $\alpha = 9/(4\pi^4)$.  In terms of the complex Dirac eigenvalues
$\lambda_n$ the partition function can also be expressed as
\begin{align}
  & Z(m_1,m_2) =
  \kakko{{\prod_n}'\mkakko{1-\frac{m_1^2}{\lambda_n^2}}
    \mkakko{1-\frac{m_2^2}{\lambda_n^2}}}_{N_f=2} \notag\\
  & = 1 - (m_1^2+m_2^2)\kakko{{\sum_n}'\frac{1}{\lambda_n^2}}
  +(m_1^4+m_2^4)\kakko{{\sum_{k<\ell}}'\frac{1}{\lambda_k^2\lambda_\ell^2}}
  \notag\\
  & + m_1^2m_2^2\kakko{\mkakko{{\sum_n}'\frac{1}{\lambda_n^2}}^2}
  - (m_1^6+m_2^6)\kakko{\sideset{}{'}\sum_{k<\ell<m}
    \frac{1}{\lambda_k^2\lambda_\ell^2\lambda_m^2}} \notag\\
  & - m_1^2m_2^2(m_1^2+m_2^2)\kakko{{\sum_n}'
    \frac{1}{\lambda_n^2}\,{\sum_{k<\ell}}'\frac{1}{\lambda_k^2\lambda_\ell^2}}
  + \mO(m^8)\,,
  \label{eq:Zmicro}
\end{align}
where the prime on the sums (and the products below) again means
$\re\lambda_n>0$.  Starting in the second line we have omitted the
subscript $N_f=2$ for brevity.  Matching of \eqref{eq:Zeff} and
\eqref{eq:Zmicro} yields massless spectral sum rules that incorporate
the effect of $\hh$, e.g.,\footnote{Similarly, we could derive
  modified massive spectral sum rules following the approach of
  \cite[sec.~4.3]{Kanazawa:2009ks}.}
\begin{align}
  \kakko{{\sum_n}' \frac{1}{\lambda_n^2}} & = - V_4\hh\,,
  \label{eq:sr1}
  \\
  \kakko{{\sum_{k<\ell}}' \frac{1}{\lambda_k^2\lambda_\ell^2}} & = \frac{1}{2}(V_4\hh)^2\,,
  \label{eq:sr2}
  \\
  \kakko{\mkakko{{\sum_n}' \frac{1}{\lambda_n^2}}^2} & = \alpha(V_4\Delta^2)^2+(V_4\hh)^2\,,
  \label{eq:sr3}
  \\
  \kakko{\sideset{}{'}\sum_{k<\ell<m} \frac{1}{\lambda_k^2\lambda_\ell^2\lambda_m^2}} 
  & = - \frac{1}{6}(V_4\hh)^3\,,
  \label{eq:sr4}
  \\
  \kakko{{\sum_n}'\frac{1}{\lambda_n^2}{\sum_{k<\ell}}' \frac{1}{\lambda_k^2\lambda_\ell^2}} 
  & = - \frac{1}{2}(V_4\hh)^3 - \alpha(V_4\Delta^2)^2(V_4\hh) \,,
  \label{eq:sr5}
\end{align}
Thus the spectral sum rules are drastically modified after the
inclusion of the $\hh$ term.  However, it is intriguing that we can
nonetheless find special combinations of sum rules that do not depend
on $\hh$. For instance, $\eqref{eq:sr3}-2\cdot\eqref{eq:sr2}$ yields
\begin{align}
  \kakko{{\sum_n}'\frac{1}{\lambda_n^4}} = \alpha(V_4\Delta^2)^2\,,
\end{align}
which agrees with the result of \cite{Kanazawa:2009ks}.  It seems
likely that there exist infinitely many such spectral sum rules that
are not affected by the $\hh$ term.  If true, the microscopic spectral
correlation functions of RMT that ignore the $\hh$ term still describe
at least some portion of the universal spectral correlations of the
small Dirac eigenvalues. In the rest of this section, we show that the
correlation functions of Dirac eigenvalues of order
$1/\sqrt{V_4\Delta^2}$ are still described by RMT and
\emph{unaffected} by the $\hh$ term.

To make sense of the modified sum rules, let us assume that the Dirac
spectrum $\{\lambda_n\}$ is a superposition of two \emph{statistically
  independent} spectra $\{a_n\}$ and $\{A_n\}$ such that
$\{\lambda_n\}=\{a_n\}\cup\{A_n\}$.  Then the expectation value in the
chiral limit $\kakko{\cdots}_{N_f}$ is understood as
\begin{align}
  \kakko{O}_{N_f} = \frac{
  {\displaystyle\int} [da][dA] ~O \mkakko{ {\prod'}_{\!\!k}a_k^2 {\prod'}_{\!\!\ell}A_\ell^2 }^{N_f}
  }{
  {\displaystyle\int} [da][dA] \mkakko{ {\prod'}_{\!\!k}a_k^2 {\prod'}_{\!\!\ell}A_\ell^2 }^{N_f}
  }\,.
\end{align}
It is essential that the measures for $\{a_n\}$ and $\{A_n\}$ are
factorized, i.e., that the two spectra are distributed independently.

Moreover, we assume that the eigenvalues $a_n$ are of order
$1/\sqrt{V_4\Delta^2}$ and satisfy the spectral sum rules
\emph{without} the $\hh$ term, i.e.,
\begin{align}
   \kakko{{\prod_n}'\mkakko{1-\frac{m_1^2}{a_n^2}}
  \mkakko{1-\frac{m_2^2}{a_n^2}}} \overset{!}{=} I_0\mkakko{\frac{3}{\pi^2}V_4\Delta^2m_1m_2},
\end{align}
whereas the $A_n$ are solely described by the $\hh$ term, i.e., 
\begin{align}
  \kakko{{\prod_n}'\mkakko{1-\frac{m_1^2}{A_n^2}}\mkakko{1-\frac{m_2^2}{A_n^2}}} 
  \overset{!}{=} \exp\big[V_4\hh(m_1^2+m_2^2)\big]\,,
  \label{eq:A_ansatz}
\end{align}
where we have again omitted the subscript $N_f=2$ for brevity.  Under
these assumptions, the mass dependence of the partition function is
reproduced correctly since
\begin{align}
  & Z(m_1,m_2) = \kakko{{\prod_n}'\mkakko{1-\frac{m_1^2}{\lambda_n^2}}
    \mkakko{1-\frac{m_2^2}{\lambda_n^2}}}
  \notag\\
  & = \bigg\langle{\prod_k}'\!\mkakko{1-\frac{m_1^2}{a_k^2}}\!
  \mkakko{1-\frac{m_2^2}{a_k^2}}
  {\prod_\ell}'\!\mkakko{1-\frac{m_1^2}{A_\ell^2}}\!
  \mkakko{1-\frac{m_2^2}{A_\ell^2}} \!\!\bigg\rangle
  \notag\\
  & = \kakko{{\prod_k}'\mkakko{1-\frac{m_1^2}{a_k^2}}
  \mkakko{1-\frac{m_2^2}{a_k^2}}
  } \notag \\
  & \quad\times \kakko{
  {\prod_\ell}'\mkakko{1-\frac{m_1^2}{A_\ell^2}}
  \mkakko{1-\frac{m_2^2}{A_\ell^2}}
  }
  \notag\\
  & = I_0\left(\frac{3}{\pi^2}V_4\Delta^2m_1m_2\right) 
  \exp\big[V_4\hh(m_1^2+m_2^2)\big]\,.
\end{align}
It follows from this that all spectral sum rules for $\{\lambda_n\}$
can be deduced from those for $\{a_n\}$ and $\{A_n\}$.  For instance,
\begin{align}
  &\kakko{\mkakko{{\sum_n}' \frac{1}{\lambda_n^2}}^2} = 
  \kakko{\mkakko{{\sum_n}' \frac{1}{a_n^2} + {\sum_n}' \frac{1}{A_n^2}}^2}
  \notag\\
  &\quad = 
  \kakko{\mkakko{{\sum_n}' \frac{1}{a_n^2}}^2} + 2
  \kakko{{\sum_n}' \frac{1}{a_n^2}} \kakko{{\sum_n}' \frac{1}{A_n^2}} 
  \notag\\
  &\quad \quad + 
  \kakko{\mkakko{{\sum_n}' \frac{1}{A_n^2}}^2} 
  \notag\\
  &\quad = \alpha(V_4\Delta^2)^2 + 2\cdot 0\cdot (-V_4\hh) + (V_4\hh)^2
  \notag\\
  &\quad = \alpha(V_4\Delta^2)^2 + (V_4\hh)^2 \,,
\end{align}
which agrees with \eqref{eq:sr3}.  The other sum rules can be
recovered similarly.

In our present argument, the spectral correlation functions obtained
from RMT are \emph{exact} results for $\{a_n\}$, but not for
$\{A_n\}$. At this stage we know nothing about $\{A_n\}$ except the
mass dependence postulated in \eqref{eq:A_ansatz}.  We now propose
that $\{A_n\}$ is the spectrum of the Dirac operator in the
\emph{free} theory, $D_\text{free}=\gamma_\nu\del_{\nu}+\mu \gamma_0$.
This may be understood as follows.  From the result obtained in
appendix \ref{app:h2} we have
\begin{align}
  \log \frac{Z(m)}{Z(0)} = V_4\big[\hh (m_1^2 + m_2^2)
  +\mO(m^4)\big]\,,
\end{align}
where there is no $\mO(m^3)$ term on the RHS because of chiral
symmetry.  In the $\epsilon$-regime at high density, where $V_4\to
\infty$ with $m\sim 1/\sqrt{V_4\Delta^2}$, the higher-order terms in
$m$ drop out, and we obtain $\log[Z(m)/Z(0)]= V_4 \hh (m_1^2 + m_2^2)$
as an exact result, i.e.,
\begin{align}
  \frac{Z(m)}{Z(0)} 
  &= {\det}'\left(1 - \frac{m_1^2}{D_\text{free}^2} \right)\left(1 -
    \frac{m_2^2}{D_\text{free}^2} \right) \notag\\
  &= \exp\big[V_4  \hh (m_1^2 + m_2^2)\big]\,.
\end{align}
Therefore the eigenvalues of the free Dirac operator satisfy
\eqref{eq:A_ansatz}, and the $\{A_n\}$ are scattered over the entire
range from $1/L$ to $\sqrt{\hh}$, where $L$ is the linear extent of
the box.  On the other hand, the typical scale of the $\{a_n\}$ is
$1/\sqrt{V_4\Delta^2}$, which is much smaller than $1/L$. This means
that the domains of $\{A_n\}$ and $\{a_n\}$ do not overlap. The
existence of the $\hh$ term does not change the spectrum on the scale
$1/\sqrt{V_4\Delta^2}$ at all, and our previous results
\cite{Kanazawa:2009en,Akemann:2010tv} based on RMT remain exact as far
as the eigenvalues on this scale are concerned.

Finally, let us recall the $\epsilon$-regime at $\mu=0$
\cite{Leutwyler:1992yt} for comparison.  There we have no $\hh
\tr(M^2)$ term because it is subleading in the
$\epsilon$-expansion. What happens if we add this term to the chiral
Lagrangian \emph{by hand}?  Then the leading spectral sum rule will be
altered to
\begin{equation}
  \kakko{ {\sum_n}' \frac{1}{\lambda_n^2} }_{\nu,N_f}
  = C_1 (V_4\Sigma)^2 + C_2 V_4 \hh\,,
  \label{eq:sum_LS_new}
\end{equation}
where $\nu$ is the topological charge, and $C_1$ and $C_2$ are
numerical constants. The second term may be interpreted as the
contribution from the perturbative Dirac spectrum since
\begin{equation}
  \kakko{{\sum_n}'\frac{1}{\lambda^2_n}}_\text{free}
  \sim \int_0^{\Lambda_\text{UV}}d\lambda\ \frac{V_4\lambda^3}{\lambda^2}
  =\frac{1}{2} V_4\Lambda^2_\text{UV}\,.
  \label{eq:free_div}
\end{equation}
However, in the limit $V_4\to\infty$ with $\Lambda_\text{UV}$ fixed,
the first term in \eqref{eq:sum_LS_new} dominates the second term, and
a conventional Leutwyler-Smilga sum rule is recovered.\footnote{It is
  essential that the cutoff is fixed in the thermodynamic limit. In
  actual lattice QCD simulations, the cutoff (the inverse lattice
  spacing) and the volume are finite, implying that the
  Leutwyler-Smilga sum rules will not be accurately satisfied.}  The
contribution of small eigenvalues dominates the UV part.

On the other hand, the contribution of small eigenvalues in dense QCD
is not large enough to overwhelm the UV part.  For example, in
\eqref{eq:sr1} the small eigenvalues contribute nothing, while the UV
spectrum yields $V_4\hh$ in the same manner as in \eqref{eq:free_div}.
In \eqref{eq:sr3} the two pieces are of the same order in the
thermodynamic limit, and one cannot drop the UV piece alone.

\section{Dirac spectrum in the free limit}
\label{app:free}

At zero chemical potential there is a marked difference between the
free and the interacting Dirac spectrum.  In the free case the
spectral density behaves like $\lambda^3$, while in the interacting
case $\rho_1(0)$ is nonzero due to the spontaneous breaking of chiral
symmetry.  To extend this comparison to nonzero chemical potential we
now compute the spectral density of the free Dirac operator at
$\mu\ne0$.  To the best of our knowledge this is a new result.

\begin{figure}
  \centering
  \includegraphics[height=60mm]{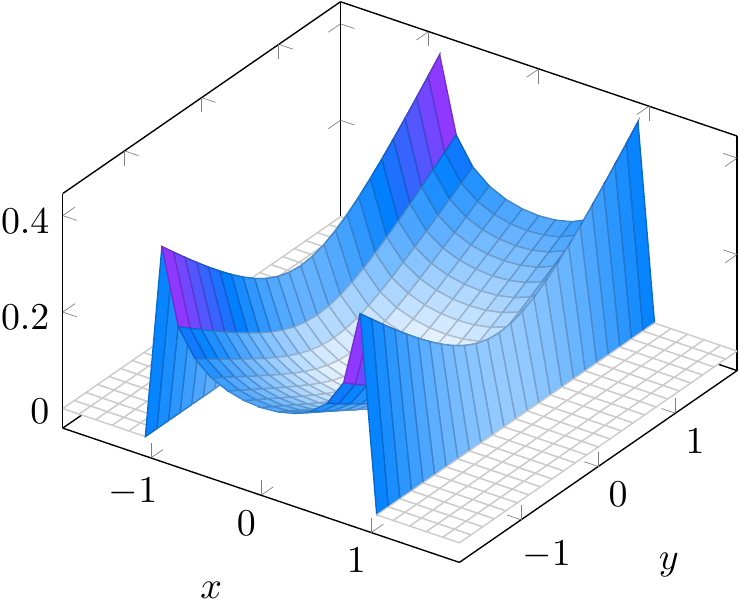}\\[3mm]
  \includegraphics[height=60mm]{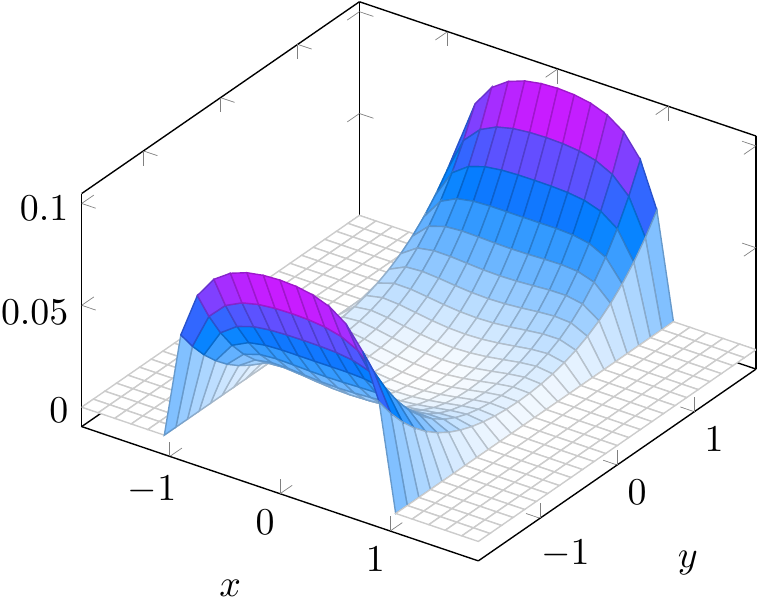}
  \caption{$\rho_1(z)$ for $d=2$ (top) and $d=4$ (bottom) at $\mu=1$.}
  \label{fg:free}
\end{figure}

The free Dirac operator in Euclidean space (where we assume even
dimension $d$) reads
\begin{align}
  \D(\mu)=(\gamma_\nu \del_\nu+\mu\gamma_0)\otimes\1_{d_\text{rep}}
  \quad\text{with}\quad \nu=1,\dots,d\,,
\end{align}
where $d_\text{rep}$ denotes the dimension of the color representation
in which the quarks transform if coupled to the gauge field.  The
spectral density of $\D(\mu)$ is defined as\footnote{Whether or not
  the factor of $1/V_d$ is included in the definition is a matter of
  convention.  Here we are consistent with \eqref{eq:one}.}
\begin{align}
  \rho_1(z=x+iy) & = \frac1{V_d}\tr\delta^2(z-D(\mu))\,,
\end{align}
where $V_d=L^d$ is the volume of a $d$-dimensional torus with linear
extent $L$.  Our main result is
\begin{align}
  \rho_1(z) &= d_\text{rep}\,\frac{2^{d/2}}{2}\,\frac{C_{d-1}}{(2\pi)^d}\,
  \frac{1}{\mu^{d-2}}\,\theta(\mu-|x|) \notag\\
  &\quad \times(x^2+y^2)\left[(\mu^2-x^2)\,(\mu^2+y^2)\right]^{\frac{d-3}{2}}\,,
  \label{eq:rho_free}
\end{align}
where $\displaystyle C_D=2\pi^{D/2}/\,\Gamma(D/2)$ is the area of a
$D$-di\-men\-sional unit sphere.  Note that \eqref{eq:rho_free} is a
continuum result and does not reproduce the spectrum of the free
lattice Dirac operator.  It is easily seen that our result has the
correct mass dimension ($d-2$) and that $\rho_1(0)=0$ in the free
case. Note also that for $|z|\ll \mu$ the effect of $\mu$ formally
drops out and $\rho_1(z)\propto |z|^2$.  Snapshots of $\rho_1(z)$ are
given for $d=2$ and $4$ in figure~\ref{fg:free} (where all parameters
were made dimensionless by an arbitrary reference scale).

Let us also check the $\mu\to 0$ limit of \eqref{eq:rho_free}. Using
\begin{align}
  \lim_{\mu\to 0}
  \frac{(\mu^2-x^2)^{\frac{d-3}{2}}}{\mu^{d-2}}\,\theta(\mu-|x|) 
  = \frac{\sqrt{\pi}\,\Gamma((d-1)/2)}{\Gamma(d/2)}\,
  \delta(x)
\end{align}
we obtain
\begin{align}
  \lim_{\mu\to 0}\rho_1(z) = 
  \frac{d_\text{rep}}{(2\pi)^{d/2}\Gamma(d/2)}\,|y|^{d-1}\,\delta(x)\,,
\end{align}
which is the correct free spectral density at $\mu=0$, see
\cite{Zyablyuk:1999aj} for $d_\text{rep}=N_c=3$ and $d=4$.

We now present the derivation of \eqref{eq:rho_free}.  Using the
relation $\bar{\del}(1/z)=\pi\delta^2(z)$ we can obtain the spectral
density from the resolvent of $\D(\mu)$, which is defined by
\begin{align}
  G(z) & = \frac1{V_d} \tr \frac{1}{z-\D(\mu)}
  \notag\\
  & = d_\text{rep}\,2^{d/2} \int
  \frac{d^dp}{(2\pi)^d}\,\frac{z}{z^2+(p_0+i\mu)^2+\bm{p}^2}\,,
\end{align}
where we have used the fact that the eigenvalues come in pairs
$\pm\lambda$, $\bm{p}$ denotes a $(d-1)$-dimensional spatial vector,
and the factor of $2^{d/2}$ is due to the $\gamma$-matrices having
dimension $2^{d/2}$. This gives
\begin{align}
  \rho_1(z) & = \frac{1}{\pi}\,\bar{\del}G(z)
  \notag\\
  & = d_\text{rep}\,\frac{2^{d/2}}{\pi} \int
  \frac{d^dp}{(2\pi)^d}\,\bar{\del}\frac{z}{z^2-w^2}\notag\\
  &\qquad\qquad\text{with } w=i\sqrt{(p_0+i\mu)^2+\bm{p}^2}
  \notag\\
  & = d_\text{rep}\,\frac{2^{d/2}}{2} \int \frac{d^dp}{(2\pi)^d}
  \big[\delta^2(z-w)+\delta^2(z+w)\big]
  \notag\\
  & = d_\text{rep}\,\frac{2^{d/2}}{2} \frac{C_{d-1}}{(2\pi)^d} \,
  \notag\\
  & \quad\times \int dp_0
  \int_0^\infty dR\,R^{d-2} \big[\delta^2(z-w)+\delta^2(z+w)\big]\,,
\end{align}
where in the last line we changed variables to spherical coordinates
with $R=|\bm{p}|$.  Writing $w=w_R+iw_I$ we have
$w_I^2-w_R^2=p_0^2-\mu^2+R^2$ and $w_Rw_I=-\mu p_0$. Therefore
\begin{align}
  R^2= w_I^2-w_R^2-p_0^2+\mu^2 =
  \frac{1}{\mu^2}(\mu^2-w_R^2)(\mu^2+w_I^2)\geq 0  
\end{align}
and thus
\begin{align}
  |w_R|\leq \mu \,.
\end{align}
The Jacobian of the transformation from $(p_0,R)$ to $(w_R,w_I)$ is
given by
\begin{align}
  \left|\frac{\del(p_0,\,R)}{\del(w_R,\,w_I)}\right| =
  \frac{|w|^2}{\mu R}\,.
\end{align}
Collecting all pieces we obtain
\begin{align}
  \rho_1(z) & = d_\text{rep}\,\frac{2^{d/2}}{2} \frac{C_{d-1}}{(2\pi)^d}
  \int_{-\mu}^\mu dw_R \notag\\
  &\quad\times\int_0^\infty dw_I \,R^{d-2} 
  \frac{|w|^2}{\mu R} \big[\delta^2(z-w)+\delta^2(z+w)\big]
  \notag\\
  & = d_\text{rep}\,\frac{2^{d/2}}{2} \frac{C_{d-1}}{(2\pi)^d}
  R^{d-3}\frac{|z|^2}{\mu} \theta(\mu-|x|) \notag\\
  &\qquad \text{with } R =
  \frac1\mu\sqrt{(\mu^2-x^2)(\mu^2+y^2)}\,,
\end{align}
which is \eqref{eq:rho_free}.

\section{Replica method}
\label{app:replica}

The purpose of this section is to present an alternative way to derive
the new Banks-Casher-type relation.  We use the so-called (fermionic)
replica method.  While this method has frequently been used in
theoretical physics over the years it can hardly be called
mathematically rigorous due to a notorious ambiguity in continuing
from integer $n$ to real $n$, where $n$ is the number of replicas. We
refer the reader to
\cite{Verbaarschot:1985qx,Verbaarschot:2005rj,Kanzieper:2009ve} for
detailed discussions of the pros and cons of the replica method. In
this section we do not discuss the validity of the replica method but
simply use it as an easy cross-check of our main result (implying that
the replica method actually works in the present case).  To add some
variety we perform the calculation for adjoint QCD rather than for
two-color QCD. Our calculation here can easily be adapted to two-color
QCD and QCD with isospin chemical potential as well.

Since the Dirac operator at nonzero $\mu$ is non-Her\-mi\-tian, we
must insert \emph{pairs of complex-conjugate replica flavors}
\cite{Girko_book,Stephanov:1996ki,Nishigaki--Kamenev}. Denoting the
number of dynamical flavors by $N_f$ and the number of pairs of
replica flavors by $n$, we have
\begin{align}
  \rho_1^{(N_f)}(z) = \frac{1}{V_4}\lim_{n\to 0}\frac{1}{\pi n}\del
  \bar\del \log Z_{N_f+2n}(z,z^*)
  \label{eq:replica_1}
\end{align}
with
\begin{align}
  & Z_{N_f+2n}(z,z^*) 
  = \kakko{{\det}^{N_f}(D+m)|\det(D+z)|^{2n}}_{\ym}
  \nonumber \\
  & = \kakko{{\det}^{N_f}(D+m){\det}^n(D+z){\det}^n(D+z^*)}_{\ym}\,,
  \label{eq:Z_adj_n}
\end{align}
where we used the property $C\gamma_5 D C\gamma_5=D^*$ of the Dirac
operator in the adjoint representation.  To evaluate
\eqref{eq:Z_adj_n} at large $\mu$ we recall the shift \eqref{eq:Eadj}
of the ground state energy density,\footnote{The possible
  $n$-dependence of $\Delta^2$ is ignored here since it causes no
  trouble as long as $\Delta^2$ is independent of $z$ and $z^*$.}
\begin{align}
  \label{eq:eadj}
  \delta\mE=-\frac{3(N_c^2-1)}{8\pi^2} \Delta^2
  & \max_{A,\Sigma_L,\Sigma_R}
  \re \Big\{A^2\tr( M\Sigma_RM^T\Sigma_L^\dag) \notag\\
  & \quad + A^{*2}\tr(M\Sigma_LM^T\Sigma_R^\dagger) \Big\}\,,
\end{align}
where the difference is that now $\Sigma_{L/R} \in
\SU(N_f+2n)/$ $\SO(N_f+2n)$ and
\begin{align}
  M = \diag(\bset{N_f}{m,\dots,m},\bset{n}{z,\dots,z},
  \bset{n}{z^*,\dots,z^*})\quad\text{with $m$ real}\,.
\end{align}
The maximum in \eqref{eq:eadj} is attained with $A = \pm1$ and
\begin{align}
  \Sigma_L = \Sigma_R = \pm \begin{pmatrix}
    \1_{N_f}&& \\ & 0&\1_n \\ &\1_n&0
  \end{pmatrix},
\end{align}
and thus $\displaystyle \delta\mE= - \frac{3(N_c^2-1)}{4\pi^2}
\Delta^2 (N_f m^2 + 2nzz^*)$.  Substituting this and $\log Z
=-V_4\delta\mE$ into \eqref{eq:replica_1} we find
\begin{align}
  \rho_1^{(N_f)}(0) = \frac{3(N_c^2-1)}{2\pi^3} \Delta^2\,.
\end{align}
This result agrees with \eqref{eq:BC_adj}.\\[1pt]
\newlength\oldparindent
\setlength\oldparindent\parindent
\begin{minipage}{\linewidth}
  \hspace*{\oldparindent}By evaluating $\delta\mE$ to higher orders in
  $M$ we could obtain a high-density analog of the Smilga-Stern
  relation \cite{Smilga:1993in}.  However, this is beyond the scope of
  this paper and deferred to future work.
\end{minipage}

\bibliographystyle{JHEP}
\bibliography{new_bc}

\end{document}